\documentclass[journal=jctcce,manuscript=article,twosides,a4paper]{achemso}
\usepackage[version=3]{mhchem} 
\usepackage[english]{babel}
\usepackage[utf8]{inputenc}
\usepackage[T1]{fontenc}
\usepackage{color}
\usepackage{rotating}
\usepackage{amssymb}
\usepackage{enumerate}
\usepackage{array,dcolumn}
\usepackage{xspace}
\usepackage{graphicx}
\usepackage{lscape}
\usepackage{rotating}
\usepackage{graphics}
\usepackage{amsmath}
\usepackage{amsfonts}
\usepackage{eso-pic}
\usepackage{epstopdf}
\usepackage{multimedia}
\usepackage{color}
\usepackage{times} 
\usepackage{bm}
\usepackage{txfonts} 
\usepackage{subcaption}

\newcommand{\ND}{\text{ND}}
\newcommand{\D}{\text{D}}

\newcommand{\dr}{^{\circ}}

\newcommand{\ie}{{\it i.e.,\ }}

\newcommand{\rrr}{\mathbf{r}}

\author{Eloy Ramos-Cordoba}
\email{eloy.raco@gmail.com}
\affiliation[Euskal Herriko Unibertsitatea]
{Kimika Fakultatea, Euskal Herriko Unibertsitatea, UPV/EHU,
and Donostia International Physics Center (DIPC). 
P.K. 1072, 20080 Donostia, Euskadi, Spain}
\alsoaffiliation{Currently at Kenneth S. Pitzer Center for Theoretical Chemistry, University of California, Berkeley}

\author{Eduard Matito}
\email{ematito@gmail.com}
\affiliation[Euskal Herriko Unibertsitatea]
{Kimika Fakultatea, Euskal Herriko Unibertsitatea, UPV/EHU,
and Donostia International Physics Center (DIPC). 
P.K. 1072, 20080 Donostia, Euskadi, Spain}
\alsoaffiliation[Ikerbasque, Basque Foundation for Science]
{IKERBASQUE, Basque Foundation for Science, 48011 Bilbao, Spain.}

\date{\today}

\title{Local Descriptors of Dynamic and \\Nondynamic Correlation}

\setkeys{acs}{articletitle = true}
\begin{document}

\begin{abstract}
Quantitatively accurate electronic structure calculations rely on the proper description
of electron correlation. A judicious choice of the approximate quantum chemistry method
depends upon the importance of dynamic and nondynamic correlation, which is usually assesed
by scalar measures. Existing measures of electron correlation do not consider separately
the regions of the Cartesian space where dynamic or nondynamic correlation are most important. 
We introduce real-space descriptors of
dynamic and nondynamic electron correlation that admit orbital decomposition.
Integration of the local descriptors yields
global numbers that can be used to quantify dynamic and nondynamic correlation.
Illustrative examples over different chemical systems with varying electron correlation
regimes are used to demonstrate the capabilities of the local descriptors.
Since the expressions only require orbitals and occupation numbers, they can be readily
applied in the context of local correlation methods, hybrid methods, density matrix
functional theory and fractional-occupancy
density functional theory.
\end{abstract}

\newpage
\section{Introduction}
A proper treatment of electron correlation is essential for an
accurate description of a quantum chemical system.
Electron correlation affects not only the electronic energy~\cite{raghavachari:91arpc,lowdin:95ijqc,raghavachari:96jpc,hattig:12cr}
but also many molecular properties~\cite{wilson:07book}
such as chemical shifts,~\cite{gauss:93jcp}
the spin,~\cite{mayer:10pccp}
optical properties,~\cite{champagne:05jcp}
the bonding character,~\cite{ruiz:16tca,matito:13pccp,matito:07fd,szalay:16arx}
or
three-center interactions.~\cite{feixas:14jctc,feixas:15ctc,rodriguez:17pccp}
The term electron correlation was introduced in quantum chemistry in 
1934 by Wigner and Seitz,~\cite{wigner:34pr} when they studied the 
cohesive energy of metals but it was L\"owdin who defined 
the electron correlation energy as the exact energy minus the Hartree-Fock (HF) 
energy.~\cite{lowdin:59acp}
L\"owdin's expression measures an electronic energy difference that reflects the
(electron correlation) effects missing in
the single-determinant HF wavefunction.
Since then, many different terms related to the electron correlation
have been coined in order to classify and recognize the deficiencies of approximate
computational methods. Hence, words such as dynamic, static, short- and long-range
correlation are now commonly employed by computational and theoretical 
chemists.~\cite{tew:07jcc,lennardjones:52jcp,callen:55jcp,hollett:11jcp,cremer:01mp,bartlett:07rcc}\newline

\vspace{-0.2cm}
A reliable description of electron correlation effects in molecular systems 
involves the use of methods that present an unfavorable scaling with the system size.
As a fitting solution,
orbital localization schemes have been widely used to bring down the computational
cost of wavefunction methods.
The use of atom-centered expansions
of many-body wavefunctions that benefit from the local character of atomic orbitals
has also become a major topic of research over the past years. In either case, functions
of localized nature are used to reduce 
the computational cost for large molecules.~\cite{zalesny:11book,ochsenfeld:07bc,ochsenfeld:04ang} 
Local descriptors of electron correlation can be likewise used to recognize the
most important interactions and regions in a molecule, 
leading to cost-efficient ways of treating electron correlation.
For instance,
local descriptors of electron correlation can be used to screen the working orbital set
or to generate a new set of orbitals 
including the most important correlation effects.
If these local descriptors
provide a separation of electron correlation effects, 
they could be also employed to combine different computational methods in the
framework of hybrid methods.~\cite{savin:88ijqc,savin:91bc,savin:96bc,leininger:97cpl}\newline

\vspace{-0.2cm}
Dynamic and nondynamic correlation (DC and NDC hereafter) 
stand out among the terms used to classify 
electron correlation effects.~\cite{sinanoglu:64acp} DC
arises from the inability of HF wavefunction to model interelectronic cusps 
and dispersion interactions, whereas NDC stems from 
near-degeneracy of HF occupied and virtual orbitals.~\cite{hollett:11jcp}
NDC is essentially a system-dependent correlation effect that calls 
for a specific treatment, while DC shows a universal 
character.~\cite{sinanoglu:64acp,lie:74jcp,valderrama:97jcp}
Many methods provide a good description of either dynamic or nondynamic 
correlation but few methods introduce both simultaneously 
in a cost-efficient manner. Therefore, it is important to recognize and quantify
the two correlation regimes. A further distinction between DC and NDC can be
made in terms of the electron density. 
DC only produces small (local) changes in the 
electron density with respect to the HF picture, 
whereas NDC induces large (global) changes in the electron density. 
The effects of electron correlation on the electron density have been used to define a
separation of the correlation energy into DC and NDC energies.~\cite{cioslowski:91pra,valderrama:97jcp,valderrama:99jcp}
We have recently used a similar strategy to separate the second-order reduced density
matrix (2-RDM) into DC and NDC parts.~\cite{ramos-cordoba:16pccp} This separation has been
applied to a two-electron model, providing
scalar descriptors of DC and NDC in 
terms of natural orbital occupancies.~\cite{ramos-cordoba:16pccp}\newline

\vspace{-0.3cm}
Many quantities have been constructed to analyze and quantify electron 
correlation~\cite{lowdin:55pr,smith:67tca,kutzelnigg:68pr,lee:89tca,lee:89ijqc,
janssen:98cpl,lee:03cpl,cioslowski:92tca,grimme:15ang,raeber:15pra,ziesche:00the,
boguslawski:12jpcl,fogueri:13tca,juhasz:06jcp,skolnik:13pra,ramos-cordoba:16pccp}
but very few provide a local account of electron correlation.
In this context, we should mention the recent paper of Grimme~\cite{grimme:15ang}
who has put forward a local NDC measure in the context of
finite-temperature density functional theory (DFT), where fractional
occupancies are used.~\cite{weinert:92prb}
In the present work we aim at introducing a local correlation function that can be 
split into DC
and NDC parts, thus providing tools for a real-space analysis of dynamic and
nondynamic correlation.
Upon integration over the whole Cartesian space, these new functions yield the 
corresponding global values recently introduced in a previous 
publication.~\cite{ramos-cordoba:16pccp}
Furthermore, these local functions 
can be decomposed into orbital contributions and they only depend
on natural orbitals and their occupancies.
To our knowledge, this is the first scheme that provides a local account of
DC and NDC contributions.
These tools can be used in the context of linear-scaling algorithms to
perform a real-space analysis of electron correlation. Since they depend on
natural orbitals and their occupancies, they can also 
serve as functional ingredients~\cite{ramos-cordoba:14jcp}
in density-matrix functional 
theories~\cite{piris:14ijqc,pernal:15tcc,cioslowski:15jcp} 
and fractional-occupancy DFT~\cite{perdew:82prl,chai:12jcp,chai:14jcp,fromager:15mp}
(or regular DFT by mapping orbital occupancies into Kohn-Sham orbital energies\cite{gruning:03jcp}).
These functions can also aid in 
designing local mixing functions~\cite{perdew:96jcp} and attenuating
functions in the context of local hybrid functionals~\cite{jaramillo:03jcp,arbuznikov:14jcp,
johnson:14jcp,silva:15jcp,henderson:07jcp,henderson:08jctc,janesko:09pccp} and range-separated
functionals.~\cite{savin:96bc,iikura:01jcp,baer:09arpc,henderson:08jpca}

\section{Theory}

We have recently put forward a natural-occupancy based index to account
for electron correlation, I$_T$, which
can be split into DC, I$_D$, and NDC, I$_{ND}$, parts~\cite{ramos-cordoba:16pccp}
\begin{eqnarray} \label{eq:it}
I_T&=& I_D + I_{\text{ND}}=\frac14\sum_{i,\sigma} [n^\sigma_i(1-n^\sigma_i)]^{1/2} \\  \label{eq:id}
I_D&=&\frac14\sum_{i,\sigma} [n^\sigma_i(1-n^\sigma_i)]^{1/2} - 2n^\sigma_i(1-n^\sigma_i) \\  
I_{\ND}&=&\frac12\sum_{i,\sigma} n^\sigma_i(1-n^\sigma_i)  \label{eq:ind}
\end{eqnarray}
where $n^\sigma_i$ is the occupation of the spin-natural orbital $i$ with spin $\sigma$.
Notice that $I_{\text{ND}}$ obtained from this derivation~\cite{ramos-cordoba:16pccp} is actually
a measure of the deviation from idempotency.~\cite{lowdin:55pr,smith:67tca}
In Ref.~\citenum{ramos-cordoba:16pccp} we suggested the decomposition of the 
2-RDM into dynamic and nondynamic parts. The application of the decomposition in the 2-RDM of 
a simple two-electron model provided the individual orbital contributions of Eq. 1-3. 
The final expressions were obtained upon the assumption that, for an $N$-electron system,
the total measure could be retrieved from 
the summation of individual orbital contributions. Therefore, the indices can be regarded as sum of orbital 
contributions to dynamic (Eq. 2) and nondynamic (Eq. 3) correlation obtained from simplified model of the 2-RDM.
By construction, these indices
can be further decomposed into orbital contributions and,
by multiplying each natural-orbital contribution by its natural orbital amplitude, $\left| \phi^\sigma_i(\rrr)\right|^2$, one can transform 
these global expressions into local functions that account for the pertinent real-space
electron correlation effects,
\begin{eqnarray}\label{eq:itr}
I_T(\rrr)&=&\frac14\sum_{i,\sigma} [n^\sigma_i(1-n^\sigma_i)]^{1/2}\left| \phi^\sigma_i(\rrr)\right|^2 \\ \label{eq:idr}
I_D(\rrr)&=&\frac14\sum_{i,\sigma} \left([n^\sigma_i(1-n^\sigma_i)]^{1/2} 
- 2n^\sigma_i(1-n^\sigma_i) \right)\left|\phi^\sigma_i(\rrr)\right|^2 \\
I_{\ND}(\rrr)&=&\frac12\sum_{i,\sigma} n^\sigma_i(1-n^\sigma_i) \left|\phi^\sigma_i(\rrr)\right|^2 \label{eq:indr}
\end{eqnarray}
where $\phi^\sigma_i$ is the spin-natural orbital $i$ with spin $\sigma$. 
Since spin-natural orbitals 
are normalized to 1, integration of the local measures (Eq.~\ref{eq:itr}-\ref{eq:indr}) over the space coordinate 
yields the global values for total, dynamic and nondynamic correlation, I$_T$, I$_D$, and I$_{ND}$, 
respectively. 
The latter expressions render themselves to straightforward orbital decompositions,
each orbital contribution being the natural orbital amplitude weighted by a
simple expression of the corresponding natural occupancy.
The simplicity of these electron correlation functions make them applicable to a wide range 
of electronic structure methods, provided a set of natural orbitals and 
occupancies are available.  

\section{Results}

\begin{figure}
\begin{center}
\includegraphics[width=0.74\textwidth]{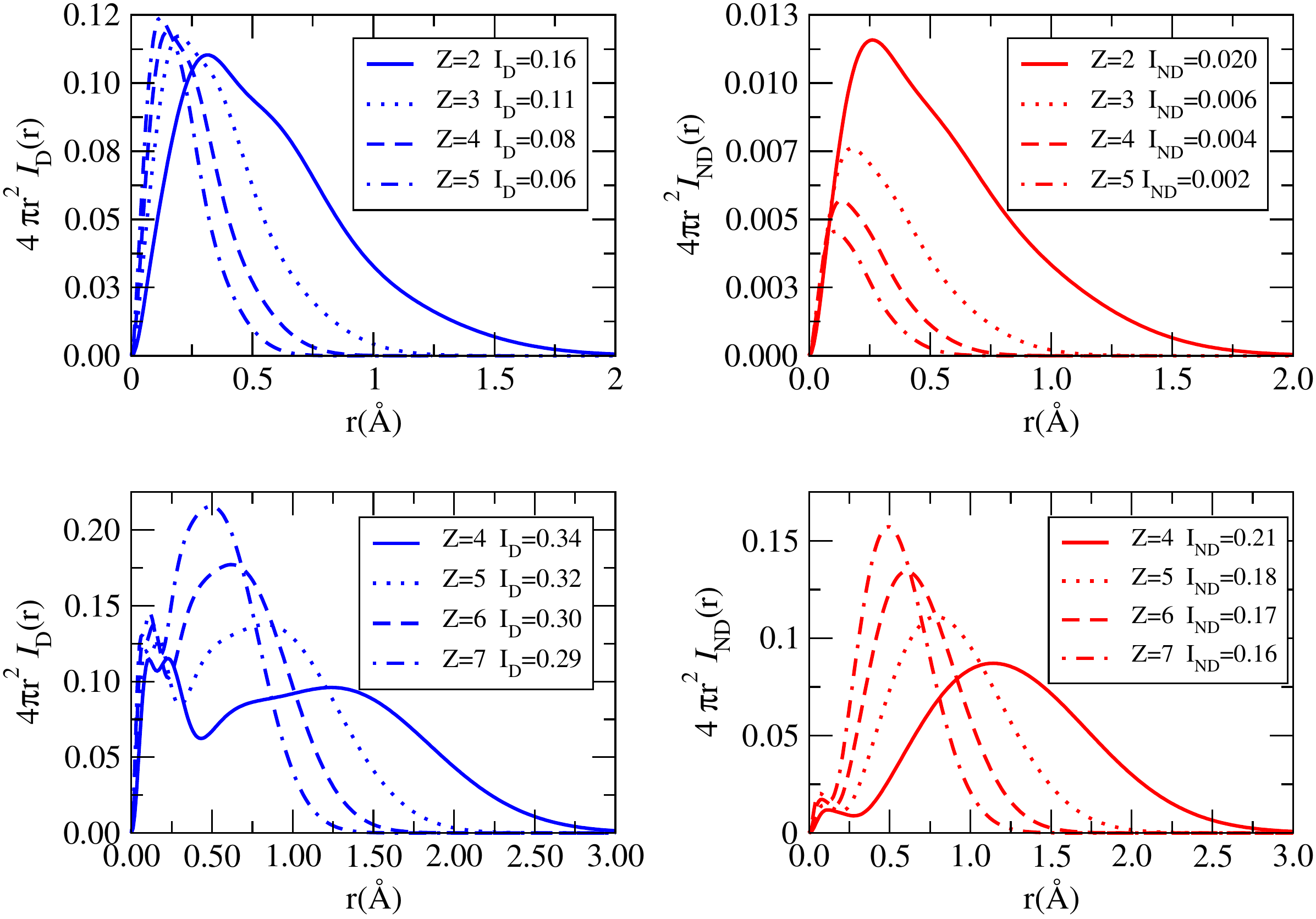}
\end{center}
\caption{Radial distributions of the 
dynamic, $I_{\D}$ (left), and nondynamic, $I_{\ND}$ (right), local functions 
for the He (top) and Be (bottom) isoelectronic series. Global values of 
$I_{\D}$ and $I_{\ND}$ are included in the legends.}
\label{f:he_series_id}
\end{figure}

\begin{figure}
\begin{center}
\includegraphics[width=0.74\textwidth]{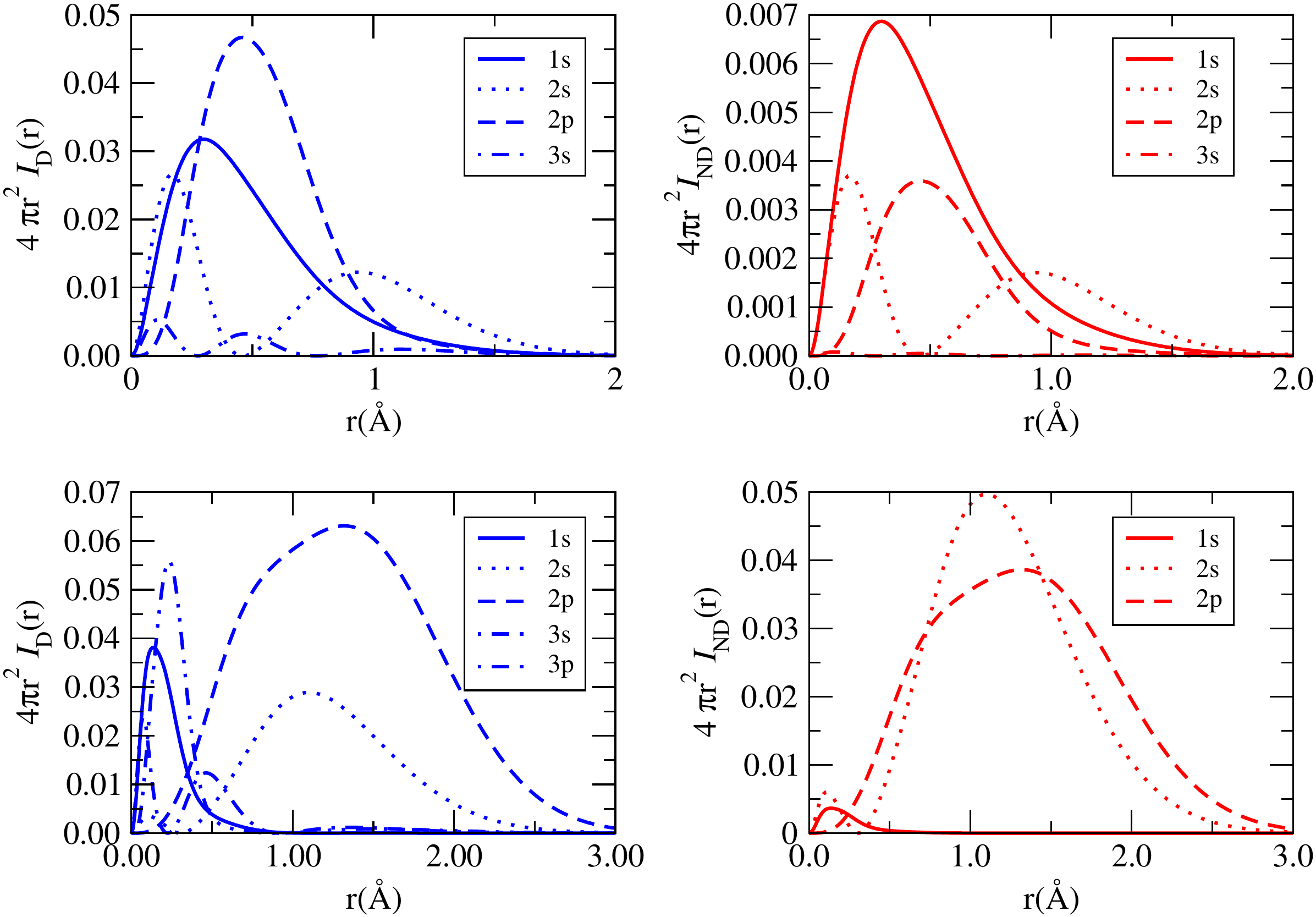}
\end{center}
\caption{Orbital contributions to the radial distributions of the
dynamic, $I_{\D}$ (left), and nondynamic, $I_{\ND}$ (right), local functions
for the He (top) and Be (bottom) atoms.}
\label{f:he_series_id_orb}
\end{figure}

All the full-configuration interaction (FCI) calculations included in this work 
are performed using a modified version of the program of Knowles and Handy,~\cite{knowles:89cpc}
whereas CASSCF and UHF calculations were obtained with Gaussian 09 package.~\cite{g09}
The local values of $I_{\D}$ and $I_{\ND}$ for the isoelectronic series of 
He and Be are shown in~\ref{f:he_series_id} in the form of the radial
distribution of such functions (\ie Eqs.~\ref{eq:idr} and~\ref{eq:indr}
integrated over the solid angle).
Since electron correlation can be tuned with $Z$, 
both systems have been previously used to study dynamic and nondynamic correlation
effects.~\cite{valderrama:97jcp,valderrama:99jcp,ramos-cordoba:16pccp} 
The He and Be series have been calculated at the FCI
level using two even-temperated basis sets optimized in a previous work~\cite{ramos-cordoba:16pccp}
using the technique described elsewhere.~\cite{matito:10pccp}
The He series is driven by dynamic 
correlation that disminishes as $Z$ increases, whereas the pseudo-degeneracy of
the $2s$ and $2p$ orbitals in Be results in important nondynamic correlation effects
in the Be isoelectronic series.
As the value of $Z$ increases the electron density shrinks as reflected by
the displacement of the local maxima of $4\pi r^2I_{\D}(r)$ and $4\pi r^2I_{\ND}(r)$
to smaller $r$ for both He and Be series. 
On the other hand, the global values, $I_{\D}$ and $I_{\ND}$, shown in the legend
of~\ref{f:he_series_id}, decrease with $Z$ because electron correlation is less important
for contracted densities.~\cite{ramos-cordoba:16pccp} 
The decomposition 
of $4\pi r^2I_{\D}(r)$ and $4\pi r^2I_{\ND}(r)$ in terms of orbital contributions 
is given in~\ref{f:he_series_id_orb}. The orbital contributions to 
$4\pi r^2I_{\D}(r)$ and $4\pi r^2I_{\ND}(r)$ for the He atom, reveal that $1s$, $2s$ 
and $2p$ orbitals are responsible for the single peak observed in the radial distribution 
function of both $I_{\D}(r)$ and $I_{\ND}(r)$. 
The Be isoelectronic series shows  
a broad peak at large $r$ for both $4\pi r^2I_{\D}(r)$ and $4\pi r^2I_{\ND}(r)$, 
and a narrower peak at small $r$ values.
The former has contributions from the $2s$ and $2p$ orbitals, while orbitals $1s$, $2s$, $3s$, and $3p$
form the peak observed in $4\pi r^2I_{\D}(r)$ at short $r$. 
The value of $r$ at which the local DC and NDC functions peak, 
$I_{\D}(r_{\text{max}})$ and $I_{\ND}(r_{\text{max}})$, 
increases with $Z$ for Be series, in line with 
the fact that the relative $2s-2p$ orbital gap decreases with $Z$.
On the other hand, in the He series $I_{\D}(r_{\text{max}})$ barely
changes with $Z$ and $I_{\ND}(r_{\text{max}})$ actually decreases with
$Z$ because NDC effects are negligible in this system.\newline

\begin{figure}[H]
\begin{center}
\includegraphics[width=0.5\textwidth]{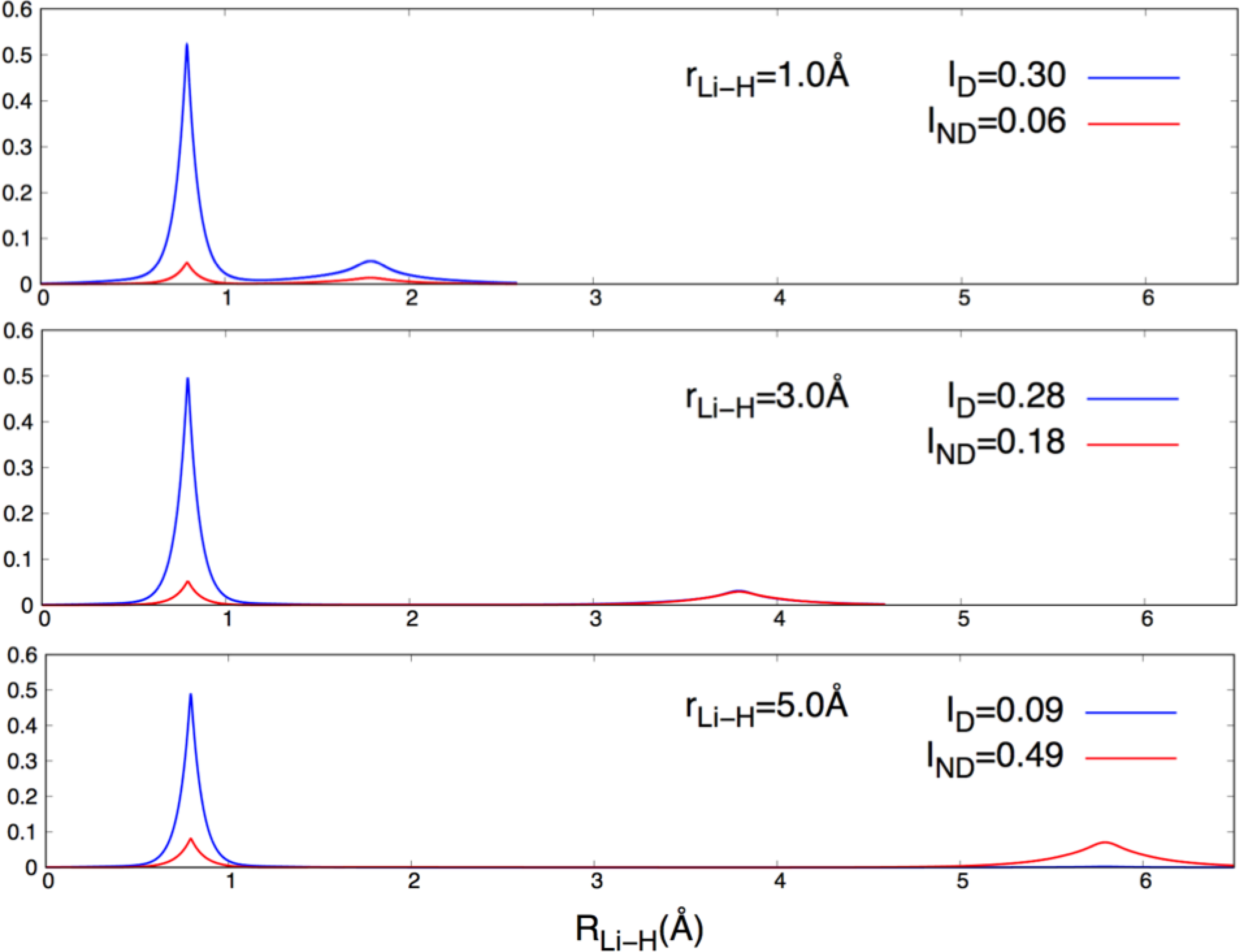}
\end{center}
\caption{Local $I_{\text{D}}$ and $I_{\text{ND}}$ along the LiH dissociation curve.}
\label{f:lih}
\end{figure}

\begin{figure}[H]
\begin{center}
\includegraphics[width=0.5\textwidth]{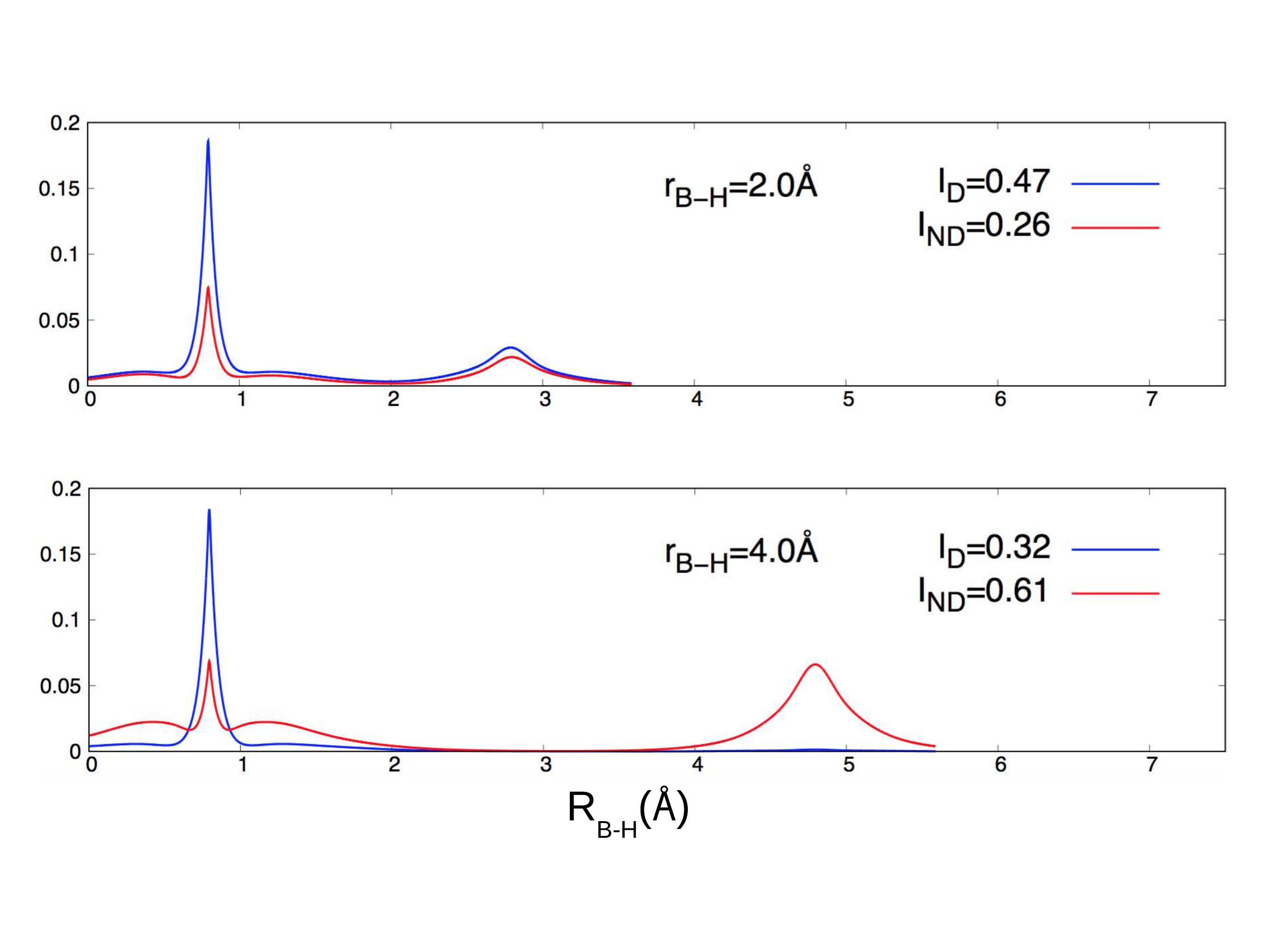}
\end{center}
\caption{Local $I_{\text{D}}$ and $I_{\text{ND}}$ along the BH dissociation curve.}	
\label{f:bh}
\end{figure}

The prototypical case of nondynamic correlation occurs in bond-breaking processes.
\ref{f:lih} and \ref{f:bh} include the plots of the local functions along the potential
energy curves of LiH and BH at the FCI level using aug-cc-pVDZ and cc-pVDZ
basis set, respectively. 
The global values of $I_{\D}$ and $I_{\ND}$ indicate
that nondynamic correlation increases and the dynamic correlation decreases as the
molecule stretches.
The local plots show that dynamic correlation is actually more important in regions 
where there is more than one localized electron, such as on Li and B atoms. As both 
systems dissociate, the $I_{\D}$ peak on H disappears because one isolated electron
cannot contribute to dynamic electron correlation.
On the other hand, 
the peak of NDC on the position of the H atom increases with $r$, a signature of the quantum
entanglement occurring between the electron in H atom and another electron of opposite spin
in the Li (or B) atom.
In the case of BH, the
NDC function presents two additional maxima at each side of the B nucleus at $r_{\text{BH}}=4.0$,
which correspond to the $2p$ orbital. 
These additional maxima increase upon dissociation 
because the $2p$ orbital that was forming the sigma bond is now singly occupied and thus NDC
predominates.

\begin{figure}[H]
\begin{center}
\includegraphics[width=0.60\textwidth]{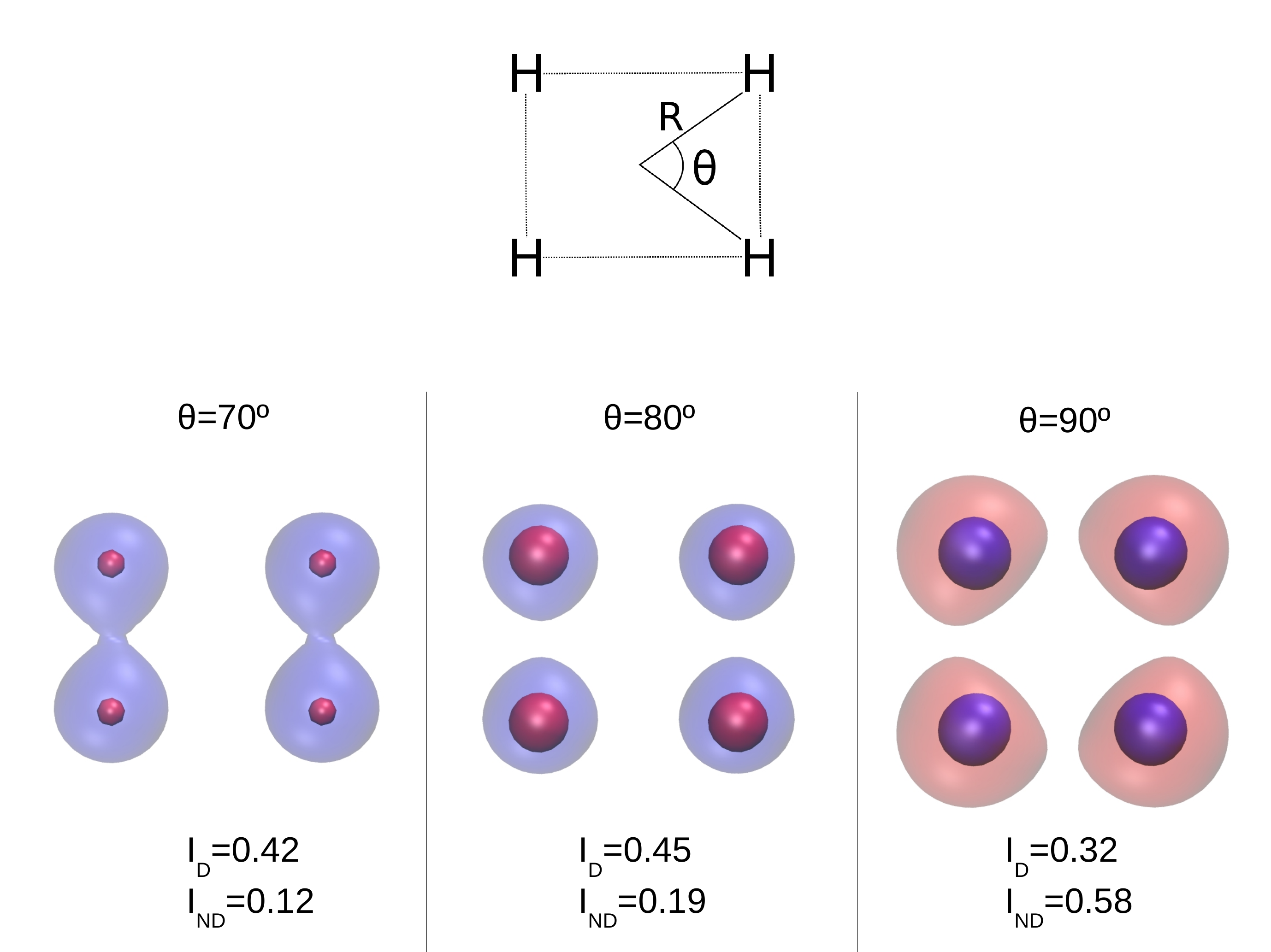}
\end{center}
\caption{Isocontour plots of $I_{\text{D}}(\rrr)$ (blue) and $I_{\text{ND}}(\rrr)$ (red) 
for H$_4$ with $R=4.0\AA$ and $\theta=70\dr$, $80\dr$, and $90\dr$. We have used an isovalue of 0.01.}
\label{f:H4}
\end{figure}

\begin{figure}[H]
\begin{center}
\includegraphics[width=0.50\textwidth]{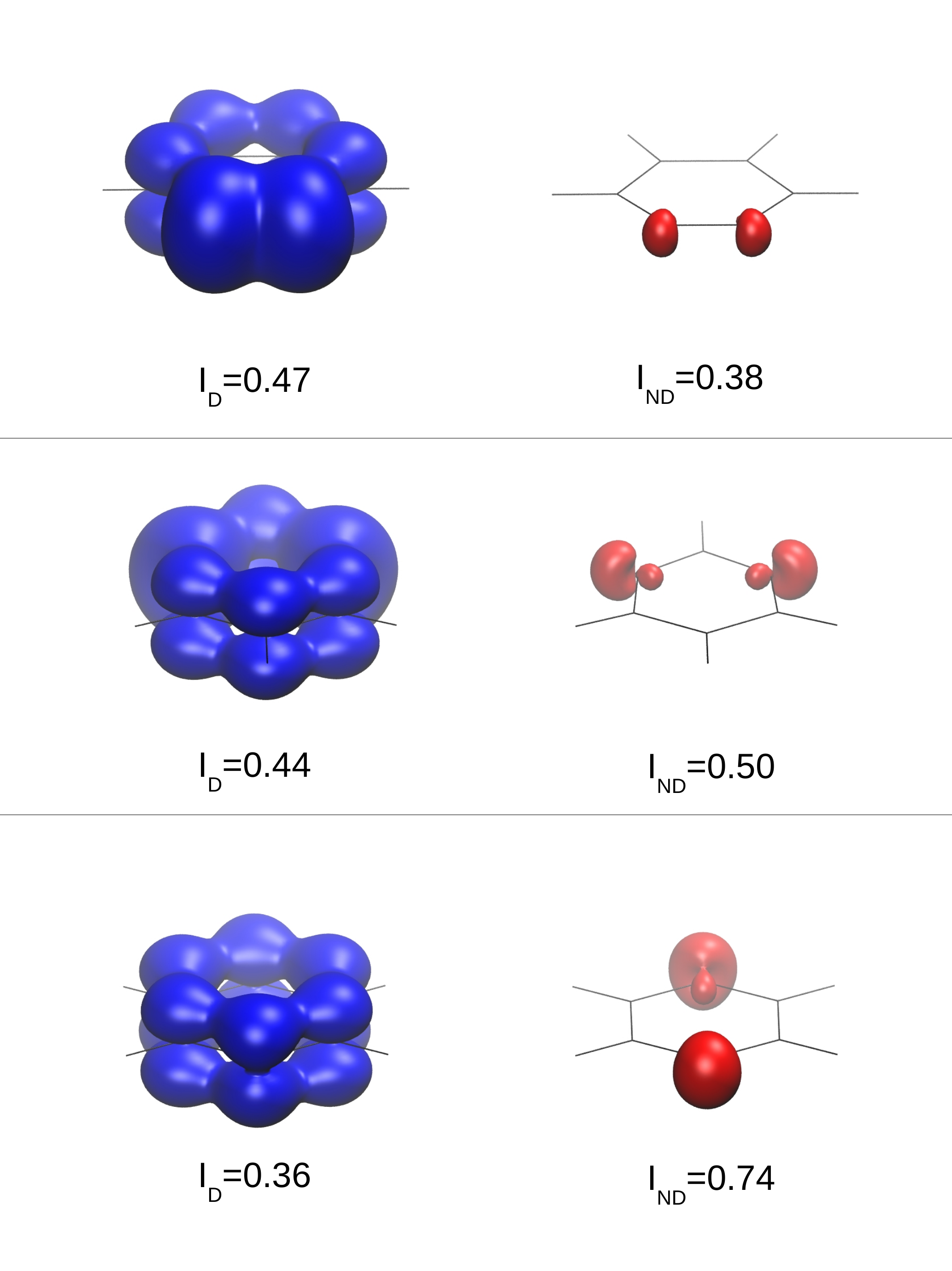}
\end{center}
\caption{ Isocontour plots of  $I_{\text{D}}(\rrr)$ (blue) and $I_{\text{ND}}(\rrr)$ (red) for ortho-, meta-, and para-benzine. We have used an isovalue of 0.01 $I_{\text{ND}}(\rrr)$ and 0.001 for $I_{\text{D}}(\rrr)$.}
\label{f:ortho_meta_para}
\end{figure}

\begin{figure}[H]
\begin{center}
\includegraphics[width=0.65\textwidth]{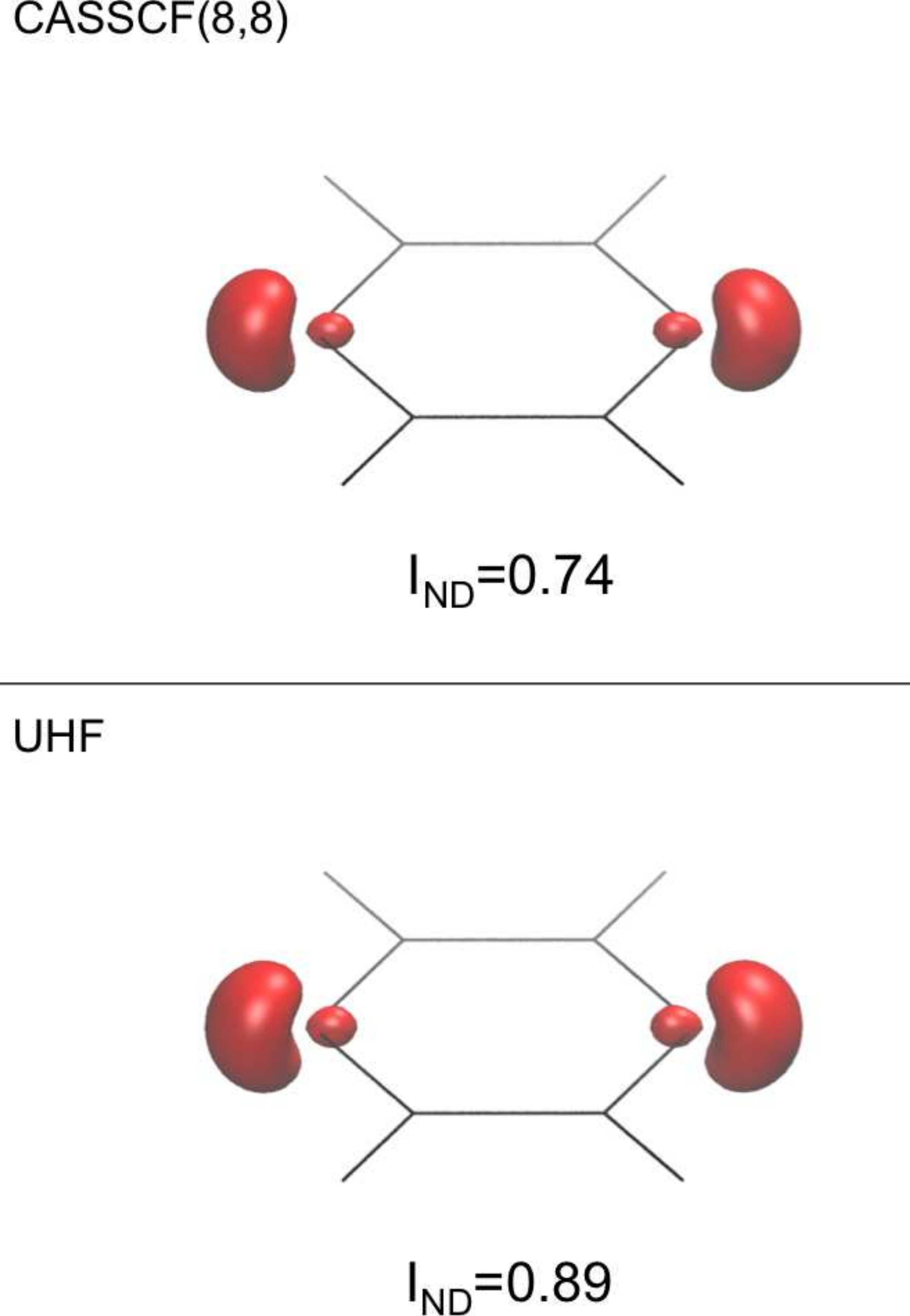}
\end{center}
\caption{Isocontour plots of $I_{\text{D}}(\rrr)$ (blue) and $I_{\text{ND}}(\rrr)$ (red) for para-benzine
for CASSCF(8,8) (top) and for UHF using UNO occupancies (bottom).
We have used an isovalue of 0.02.}
\label{f:para_cas_uhf}
\end{figure}

D$_{2h}$/D$_{4h}$ potential energy surface (PES) of H$_4$ has been used to
assess the accuracy of electronic structure methods to account for simultaneous DC and 
NDC.~\cite{ramos-cordoba:15jcp,bulik:15jctc,robinson:12jcp,vanvoorhis:00jcp,jankowski:99jcp}
The PES, calculated at the FCI level using aug-cc-pVDZ basis set,
is described 
by two parameters, $R$ and $\theta$ (see \ref{f:H4}), $\theta\approx90\dr$ and large $R$
corresponding to the structures most affected by NDC.
In~\ref{f:H4} we find the isosurface plots of
$I_{\text{D}}(\rrr)$ and $I_{\text{ND}}(\rrr)$ for 
$\theta=70\dr$, $80\dr$, and $90\dr$, keeping R fixed at 0.80$\AA$. From $\theta=70\dr$ 
to $\theta=90\dr$ there is an increase of NDC due the degeneracy of b$_{3u}$ and b$_{2u}$
orbitals,~\cite{ramos-cordoba:15jcp} as confirmed by the red isosurfaces in \ref{f:H4}.
At $\theta=70\dr$, $I_{\text{D}}(\rrr)$ extends along two H-H bonds while $I_{\text{ND}}(\rrr)$ 
is almost negligible. The $I_{\text{ND}}(\rrr)$ increases as we move towards the 
$D_{4h}$ geometry, becoming larger than $I_{\text{D}}(\rrr)$ at $\theta=90\dr$. The local
descriptors thus provide an appropriate real-space account of DC and NDC in this system.
Finally, we have also have investigated ortho-, meta- and para-benzine singlet diradicals,
which have two unpaired
electrons and, therefore, are affected by nondynamic correlation. An effective nondynamic
localization function should identify the regions where these unpaired electrons localize,
which coincide with the regions where the local spin~\cite{ramos-cordoba:12jctc,ramos-cordoba:12pccp}
is larger.~\cite{ramos-cordoba:14pccp}
Natural orbitals and occupations were calculated from a CASSCF/cc-pVTZ wavefunction
with 8 active electrons in 8 orbitals including six $\pi$-orbitals and two $\sigma$-
orbitals at the geometries of the given level of theory.
NDC arises in these systems due to the near-degeneracy of
the HOMO and LUMO $\sigma$-orbitals, which are localized at the position of the radical centers.
The three isomers have an $I_{\text{ND}}(\rrr)$ localized at the radical positions, showing the regions of the 
molecule where NDC is important (see~\ref{f:ortho_meta_para}). $I_{\text{D}}(\rrr)$ is delocalized 
along the $\pi-$orbitals and only has some small contributions from the $\sigma$-orbitals 
for ortho- and meta-benzine. 
To illustrate that the local correlation functions introduced in this work can 
be applied to any electronic structure method providing a set of orbitals and
fractional occupations, we have computed 
$I_{\text{D}}(\rrr)$ and $I_{\text{ND}}(\rrr)$ for para-benzine 
at the UHF/cc-pVTZ level.
Some NDC can be captured in a HF wavefunction by breaking the spin symmetry
at the expense of not being an eigenfunction of $  \hat S^2 $.
The unrestricted natural orbitals (UNO)~\cite{pulay:88jcp,harriman:64jcp}
are often used to generate starting orbitals for MCSCF/CASSF wavefunctions 
computed in regions of the potential energy surface where nondynamic electron correlation 
is dominant. The UNOs are obtained by diagonalization of the transformed
UHF charge density matrix.~\cite{pulay:88jcp} The UNO occupancies can be input in Eq.~\ref{eq:ind} and~\ref{eq:indr}
to assess the global and local nondynamic correlation obtained from the UHF wavefunction.
 We can observe in \ref{f:para_cas_uhf} that 
$I_{\text{ND}}(\rrr)$ isosurfaces are qualitatively equivalent for UNO and CASSCF(8,8)
natural orbitals indicating than both methods, UHF and CASSCF,
capture NDC in the region where the unpaired electrons are localized in the molecule.

In principle, the UNOs and their occupancies can also be used to compute $I_{\text{D}}(r)$ (Eq.\ref{eq:idr}) 
and $I_{\text{D}}$ (Eq.\ref{eq:id}). 
The total value of $I_{\text{D}}$ for para-benzine at the UHF level is 0.60, which is quite high for the UHF 
wavefunction that should not contain dynamic correlation. However, it is important to note that UNOs 
are the natural orbitals of the spin-projected UHF wave function and, as such, they should only be used 
to qualitatively understand nondynamic electron correlation.\cite{pulay:88jcp,harriman:64jcp}
In our opinion, it is unlikely that UNOs reflect the dynamic correlation contents of the UHF.

\section{Conclusions}

In this paper we have introduced real-space descriptors of dynamic and nondynamic electron correlation.
Theses local functions
can be constructed from any electronic structure calculation provided a set of orbitals and
occupation numbers are available. The functions admit orbital decomposition
and, upon integration over the Cartesian space, yield
the corresponding scalar descriptors recently proposed in 
Ref~\citenum{ramos-cordoba:16pccp}. 
The local correlation functions
have been computed for a series of systems with varying electron correlation regimes,
showing their ability to differentiate regions 
of the molecule where dynamic or nondynamic correlation are important. 
The DC and NDC local functions presented in this
work hold the promise to aid in the development of new local correlation methods
and hybrid methods.

\section*{Acknowledgements}
The authors thank Prof. Lude\~na and Prof. Salvador for insightful discussions.
This research has been funded by Spanish MINECO/FEDER
Project No. CTQ2014-52525-P and the Basque 
Country Consolidated Group Project No. IT588-13.
Technical and human support provided by IZO-SGI, SGIker (UPV/EHU, 
MICINN, GV/EJ, ERDF and ESF) is gratefully acknowledged. 
ERC acknowledges funding from the European Union’s
Horizon 2020 research and innovation programme under the
Marie Sklodowska-Curie grant agreement (No. 660943).
\bibliography{gen}

\providecommand{\latin}[1]{#1}
\makeatletter
\providecommand{\doi}
  {\begingroup\let\do\@makeother\dospecials
  \catcode`\{=1 \catcode`\}=2 \doi@aux}
\providecommand{\doi@aux}[1]{\endgroup\texttt{#1}}
\makeatother
\providecommand*\mcitethebibliography{\thebibliography}
\csname @ifundefined\endcsname{endmcitethebibliography}
  {\let\endmcitethebibliography\endthebibliography}{}
\begin{mcitethebibliography}{85}
\providecommand*\natexlab[1]{#1}
\providecommand*\mciteSetBstSublistMode[1]{}
\providecommand*\mciteSetBstMaxWidthForm[2]{}
\providecommand*\mciteBstWouldAddEndPuncttrue
  {\def\EndOfBibitem{\unskip.}}
\providecommand*\mciteBstWouldAddEndPunctfalse
  {\let\EndOfBibitem\relax}
\providecommand*\mciteSetBstMidEndSepPunct[3]{}
\providecommand*\mciteSetBstSublistLabelBeginEnd[3]{}
\providecommand*\EndOfBibitem{}
\mciteSetBstSublistMode{f}
\mciteSetBstMaxWidthForm{subitem}{(\alph{mcitesubitemcount})}
\mciteSetBstSublistLabelBeginEnd
  {\mcitemaxwidthsubitemform\space}
  {\relax}
  {\relax}

\bibitem[Raghavachari(1991)]{raghavachari:91arpc}
Raghavachari,~K. Electron correlation techniques in quantum chemistry: Recent
  advances. \emph{Ann. Rev. Phys. Chem.} \textbf{1991}, \emph{42},
  615--642\relax
\mciteBstWouldAddEndPuncttrue
\mciteSetBstMidEndSepPunct{\mcitedefaultmidpunct}
{\mcitedefaultendpunct}{\mcitedefaultseppunct}\relax
\EndOfBibitem
\bibitem[L{\"o}wdin(1995)]{lowdin:95ijqc}
L{\"o}wdin,~P.-O. The historical development of the electron correlation
  problem. \emph{Int. J. Quant. Chem.} \textbf{1995}, \emph{55}, 77--102\relax
\mciteBstWouldAddEndPuncttrue
\mciteSetBstMidEndSepPunct{\mcitedefaultmidpunct}
{\mcitedefaultendpunct}{\mcitedefaultseppunct}\relax
\EndOfBibitem
\bibitem[Raghavachari and Anderson(1996)Raghavachari, and
  Anderson]{raghavachari:96jpc}
Raghavachari,~K.; Anderson,~J.~B. Electron correlation effects in molecules.
  \emph{J. Phys. Chem.} \textbf{1996}, \emph{100}, 12960--12973\relax
\mciteBstWouldAddEndPuncttrue
\mciteSetBstMidEndSepPunct{\mcitedefaultmidpunct}
{\mcitedefaultendpunct}{\mcitedefaultseppunct}\relax
\EndOfBibitem
\bibitem[H{\"a}ttig \latin{et~al.}(2012)H{\"a}ttig, Klopper, K{\"o}hn, and
  Tew]{hattig:12cr}
H{\"a}ttig,~C.; Klopper,~W.; K{\"o}hn,~A.; Tew,~D.~P. Explicitly correlated
  electrons in molecules. \emph{Chem. Rev.} \textbf{2012}, \emph{112},
  4--74\relax
\mciteBstWouldAddEndPuncttrue
\mciteSetBstMidEndSepPunct{\mcitedefaultmidpunct}
{\mcitedefaultendpunct}{\mcitedefaultseppunct}\relax
\EndOfBibitem
\bibitem[Wilson(2007)]{wilson:07book}
Wilson,~S. \emph{Electron correlation in molecules}; Dover: New York,
  2007\relax
\mciteBstWouldAddEndPuncttrue
\mciteSetBstMidEndSepPunct{\mcitedefaultmidpunct}
{\mcitedefaultendpunct}{\mcitedefaultseppunct}\relax
\EndOfBibitem
\bibitem[Gauss(1993)]{gauss:93jcp}
Gauss,~J. Effects of electron correlation in the calculation of nuclear
  magnetic resonance chemical shifts. \emph{J. Chem. Phys.} \textbf{1993},
  \emph{99}, 3629--3643\relax
\mciteBstWouldAddEndPuncttrue
\mciteSetBstMidEndSepPunct{\mcitedefaultmidpunct}
{\mcitedefaultendpunct}{\mcitedefaultseppunct}\relax
\EndOfBibitem
\bibitem[Mayer and Matito(2010)Mayer, and Matito]{mayer:10pccp}
Mayer,~I.; Matito,~E. Calculation of local spins for correlated wave functions.
  \emph{Phys. Chem. Chem. Phys.} \textbf{2010}, \emph{10}, 11308--11314\relax
\mciteBstWouldAddEndPuncttrue
\mciteSetBstMidEndSepPunct{\mcitedefaultmidpunct}
{\mcitedefaultendpunct}{\mcitedefaultseppunct}\relax
\EndOfBibitem
\bibitem[Champagne \latin{et~al.}(2005)Champagne, Botek, Nakano, Nitta, and
  Yamaguchi]{champagne:05jcp}
Champagne,~B.; Botek,~E.; Nakano,~M.; Nitta,~T.; Yamaguchi,~K. Basis set and
  electron correlation effects on the polarizability and second
  hyperpolarizability of model open-shell $\pi$-conjugated systems. \emph{J.
  Chem. Phys.} \textbf{2005}, \emph{122}, 114315\relax
\mciteBstWouldAddEndPuncttrue
\mciteSetBstMidEndSepPunct{\mcitedefaultmidpunct}
{\mcitedefaultendpunct}{\mcitedefaultseppunct}\relax
\EndOfBibitem
\bibitem[Ruiz \latin{et~al.}(2016)Ruiz, Matito, Holgu{\'\i}n-Gallego,
  Francisco, Pend{\'a}s, and Rocha-Rinza]{ruiz:16tca}
Ruiz,~I.; Matito,~E.; Holgu{\'\i}n-Gallego,~F.~J.; Francisco,~E.;
  Pend{\'a}s,~{\'A}.~M.; Rocha-Rinza,~T. Fermi and Coulomb correlation effects
  upon the interacting quantum atoms energy partition. \emph{Theor. Chem. Acc.}
  \textbf{2016}, \emph{135}, 209\relax
\mciteBstWouldAddEndPuncttrue
\mciteSetBstMidEndSepPunct{\mcitedefaultmidpunct}
{\mcitedefaultendpunct}{\mcitedefaultseppunct}\relax
\EndOfBibitem
\bibitem[Matito \latin{et~al.}(2013)Matito, Salvador, and
  Styszy{\'n}ski]{matito:13pccp}
Matito,~E.; Salvador,~P.; Styszy{\'n}ski,~J. Benchmark calculations of metal
  carbonyl cations: relativistic vs. electron correlation effects. \emph{Phys.
  Chem. Chem. Phys.} \textbf{2013}, \emph{15}, 20080--20090\relax
\mciteBstWouldAddEndPuncttrue
\mciteSetBstMidEndSepPunct{\mcitedefaultmidpunct}
{\mcitedefaultendpunct}{\mcitedefaultseppunct}\relax
\EndOfBibitem
\bibitem[Matito \latin{et~al.}(2007)Matito, Sol\`a, Salvador, and
  Duran]{matito:07fd}
Matito,~E.; Sol\`a,~M.; Salvador,~P.; Duran,~M. Electron sharing indexes at the
  correlated level. Application to aromaticity calculations. \emph{Faraday
  Discuss.} \textbf{2007}, \emph{135}, 325--345\relax
\mciteBstWouldAddEndPuncttrue
\mciteSetBstMidEndSepPunct{\mcitedefaultmidpunct}
{\mcitedefaultendpunct}{\mcitedefaultseppunct}\relax
\EndOfBibitem
\bibitem[Szalay \latin{et~al.}(2016)Szalay, Barcza, Szilv{\'a}si, Veis, and
  Legeza]{szalay:16arx}
Szalay,~S.; Barcza,~G.; Szilv{\'a}si,~T.; Veis,~L.; Legeza,~{\"O}. The
  correlation theory of the chemical bond. \emph{arXiv preprint
  arXiv:1605.06919} \textbf{2016}, \relax
\mciteBstWouldAddEndPunctfalse
\mciteSetBstMidEndSepPunct{\mcitedefaultmidpunct}
{}{\mcitedefaultseppunct}\relax
\EndOfBibitem
\bibitem[Feixas \latin{et~al.}(2014)Feixas, Sol{\`a}, Barroso, Ugalde, and
  Matito]{feixas:14jctc}
Feixas,~F.; Sol{\`a},~M.; Barroso,~J.~M.; Ugalde,~J.~M.; Matito,~E. New
  Approximation to the Third-Order Density. Application to the Calculation of
  Correlated Multicenter Indices. \emph{J. Chem. Theory Comput.} \textbf{2014},
  \emph{10}, 3055--3065\relax
\mciteBstWouldAddEndPuncttrue
\mciteSetBstMidEndSepPunct{\mcitedefaultmidpunct}
{\mcitedefaultendpunct}{\mcitedefaultseppunct}\relax
\EndOfBibitem
\bibitem[Feixas \latin{et~al.}(2015)Feixas, Rodr{\'\i}guez-Mayorga, Matito, and
  Sol{\`a}]{feixas:15ctc}
Feixas,~F.; Rodr{\'\i}guez-Mayorga,~M.; Matito,~E.; Sol{\`a},~M. Three-center
  bonding analyzed from correlated and uncorrelated third-order reduced density
  matrices. \emph{Comput. Theor. Chem.} \textbf{2015}, \emph{1053},
  173--179\relax
\mciteBstWouldAddEndPuncttrue
\mciteSetBstMidEndSepPunct{\mcitedefaultmidpunct}
{\mcitedefaultendpunct}{\mcitedefaultseppunct}\relax
\EndOfBibitem
\bibitem[Wigner and Seitz(1934)Wigner, and Seitz]{wigner:34pr}
Wigner,~E.; Seitz,~F. On the constitution of metallic sodium. II. \emph{Phys.
  Rev.} \textbf{1934}, \emph{46}, 509\relax
\mciteBstWouldAddEndPuncttrue
\mciteSetBstMidEndSepPunct{\mcitedefaultmidpunct}
{\mcitedefaultendpunct}{\mcitedefaultseppunct}\relax
\EndOfBibitem
\bibitem[L{\"o}wdin(1959)]{lowdin:59acp}
L{\"o}wdin,~P.-O. Correlation Problem in Many-Electron Quantum Mechanics I.
  Review of Different Approaches and Discussion of Some Current Ideas.
  \emph{Adv. Chem. Phys.} \textbf{1959}, \emph{2}, 207--322\relax
\mciteBstWouldAddEndPuncttrue
\mciteSetBstMidEndSepPunct{\mcitedefaultmidpunct}
{\mcitedefaultendpunct}{\mcitedefaultseppunct}\relax
\EndOfBibitem
\bibitem[Tew \latin{et~al.}(2007)Tew, Klopper, and Helgaker]{tew:07jcc}
Tew,~D.~P.; Klopper,~W.; Helgaker,~T. Electron correlation: The many-body
  problem at the heart of chemistry. \emph{J. Comput. Chem.} \textbf{2007},
  \emph{28}, 1307--1320\relax
\mciteBstWouldAddEndPuncttrue
\mciteSetBstMidEndSepPunct{\mcitedefaultmidpunct}
{\mcitedefaultendpunct}{\mcitedefaultseppunct}\relax
\EndOfBibitem
\bibitem[Lennard-Jones(1952)]{lennardjones:52jcp}
Lennard-Jones,~J.~E. The spatial correlation of electrons in molecules.
  \emph{J. Chem. Phys.} \textbf{1952}, \emph{20}, 1024\relax
\mciteBstWouldAddEndPuncttrue
\mciteSetBstMidEndSepPunct{\mcitedefaultmidpunct}
{\mcitedefaultendpunct}{\mcitedefaultseppunct}\relax
\EndOfBibitem
\bibitem[Callen(1955)]{callen:55jcp}
Callen,~E. Configuration interaction applied to the hydrogen molecule. \emph{J.
  Chem. Phys.} \textbf{1955}, \emph{23}, 360--362\relax
\mciteBstWouldAddEndPuncttrue
\mciteSetBstMidEndSepPunct{\mcitedefaultmidpunct}
{\mcitedefaultendpunct}{\mcitedefaultseppunct}\relax
\EndOfBibitem
\bibitem[Hollett and Gill(2011)Hollett, and Gill]{hollett:11jcp}
Hollett,~J.~W.; Gill,~P.~M. The two faces of static correlation. \emph{J. Chem.
  Phys.} \textbf{2011}, \emph{134}, 114111\relax
\mciteBstWouldAddEndPuncttrue
\mciteSetBstMidEndSepPunct{\mcitedefaultmidpunct}
{\mcitedefaultendpunct}{\mcitedefaultseppunct}\relax
\EndOfBibitem
\bibitem[Cremer(2001)]{cremer:01mp}
Cremer,~D. Density functional theory: coverage of dynamic and non-dynamic
  electron correlation effects. \emph{Molec. Phys.} \textbf{2001}, \emph{99},
  1899--1940\relax
\mciteBstWouldAddEndPuncttrue
\mciteSetBstMidEndSepPunct{\mcitedefaultmidpunct}
{\mcitedefaultendpunct}{\mcitedefaultseppunct}\relax
\EndOfBibitem
\bibitem[Bartlett and Stanton(2007)Bartlett, and Stanton]{bartlett:07rcc}
Bartlett,~R.~J.; Stanton,~J.~F. Applications of Post-Hartree-Fock Methods: A
  Tutorial. \emph{Rev. Comp. Chem.} \textbf{2007}, \emph{5}, 65--169\relax
\mciteBstWouldAddEndPuncttrue
\mciteSetBstMidEndSepPunct{\mcitedefaultmidpunct}
{\mcitedefaultendpunct}{\mcitedefaultseppunct}\relax
\EndOfBibitem
\bibitem[Zalesny \latin{et~al.}(2011)Zalesny, Papadopoulos, Mezey, and
  Leszczynski]{zalesny:11book}
Zalesny,~R.; Papadopoulos,~M.~G.; Mezey,~P.~G.; Leszczynski,~J.
  \emph{Linear-scaling techniques in computational chemistry and physics:
  Methods and applications}; Springer Science+ Business Media BV, 2011\relax
\mciteBstWouldAddEndPuncttrue
\mciteSetBstMidEndSepPunct{\mcitedefaultmidpunct}
{\mcitedefaultendpunct}{\mcitedefaultseppunct}\relax
\EndOfBibitem
\bibitem[Ochsenfeld \latin{et~al.}(2007)Ochsenfeld, Kussmann, and
  Lambrecht]{ochsenfeld:07bc}
Ochsenfeld,~C.; Kussmann,~J.; Lambrecht,~D.~S. \emph{Reviews in Computational
  Chemistry}; John Wiley \& Sons, Inc., 2007; pp 1--82\relax
\mciteBstWouldAddEndPuncttrue
\mciteSetBstMidEndSepPunct{\mcitedefaultmidpunct}
{\mcitedefaultendpunct}{\mcitedefaultseppunct}\relax
\EndOfBibitem
\bibitem[Ochsenfeld \latin{et~al.}(2004)Ochsenfeld, Kussmann, and
  Koziol]{ochsenfeld:04ang}
Ochsenfeld,~C.; Kussmann,~J.; Koziol,~F. Ab Initio NMR Spectra for Molecular
  Systems with a Thousand and More Atoms: A Linear-Scaling Method. \emph{Angew.
  Chem. Int. Ed. Engl.} \textbf{2004}, \emph{43}, 4485--4489\relax
\mciteBstWouldAddEndPuncttrue
\mciteSetBstMidEndSepPunct{\mcitedefaultmidpunct}
{\mcitedefaultendpunct}{\mcitedefaultseppunct}\relax
\EndOfBibitem
\bibitem[Savin(1988)]{savin:88ijqc}
Savin,~A. A combined density functional and configuration interaction method.
  \emph{Int. J. Quant. Chem.} \textbf{1988}, \emph{34}, 59--69\relax
\mciteBstWouldAddEndPuncttrue
\mciteSetBstMidEndSepPunct{\mcitedefaultmidpunct}
{\mcitedefaultendpunct}{\mcitedefaultseppunct}\relax
\EndOfBibitem
\bibitem[Savin(1991)]{savin:91bc}
Savin,~A. \emph{Density functional methods in chemistry}; Springer, 1991; pp
  213--230\relax
\mciteBstWouldAddEndPuncttrue
\mciteSetBstMidEndSepPunct{\mcitedefaultmidpunct}
{\mcitedefaultendpunct}{\mcitedefaultseppunct}\relax
\EndOfBibitem
\bibitem[Savin(1996)]{savin:96bc}
Savin,~A. In \emph{Recent Developments of Modern Density Functional Theory};
  Seminario,~J.~M., Ed.; Elsevier: Amsterdam, 1996; p 327\relax
\mciteBstWouldAddEndPuncttrue
\mciteSetBstMidEndSepPunct{\mcitedefaultmidpunct}
{\mcitedefaultendpunct}{\mcitedefaultseppunct}\relax
\EndOfBibitem
\bibitem[Leininger \latin{et~al.}(1997)Leininger, Stoll, Werner, and
  Savin]{leininger:97cpl}
Leininger,~T.; Stoll,~H.; Werner,~H.-J.; Savin,~A. Combining long-range
  configuration interaction with short-range density functionals. \emph{Chem.
  Phys. Lett.} \textbf{1997}, \emph{275}, 151--160\relax
\mciteBstWouldAddEndPuncttrue
\mciteSetBstMidEndSepPunct{\mcitedefaultmidpunct}
{\mcitedefaultendpunct}{\mcitedefaultseppunct}\relax
\EndOfBibitem
\bibitem[Sinano{\u{g}}lu(1964)]{sinanoglu:64acp}
Sinano{\u{g}}lu,~O. Many-Electron Theory of Atoms, Molecules and Their
  Interactions. \emph{Adv. Chem. Phys.} \textbf{1964}, \emph{6}, 315--412\relax
\mciteBstWouldAddEndPuncttrue
\mciteSetBstMidEndSepPunct{\mcitedefaultmidpunct}
{\mcitedefaultendpunct}{\mcitedefaultseppunct}\relax
\EndOfBibitem
\bibitem[Lie and Clementi(1974)Lie, and Clementi]{lie:74jcp}
Lie,~G.~C.; Clementi,~E. Study of the electronic structure of molecules. XXI.
  Correlation energy corrections as a functional of the Hartree-Fock density
  and its application to the hydrides of the second row atoms. \emph{J. Chem.
  Phys.} \textbf{1974}, \emph{60}, 1275--1287\relax
\mciteBstWouldAddEndPuncttrue
\mciteSetBstMidEndSepPunct{\mcitedefaultmidpunct}
{\mcitedefaultendpunct}{\mcitedefaultseppunct}\relax
\EndOfBibitem
\bibitem[Valderrama \latin{et~al.}(1997)Valderrama, Lude{\~n}a, and
  Hinze]{valderrama:97jcp}
Valderrama,~E.; Lude{\~n}a,~E.~V.; Hinze,~J. {Analysis of dynamical and
  nondynamical components of electron correlation energy by means of
  local-scaling density-functional theory}. \emph{J. Chem. Phys.}
  \textbf{1997}, \emph{106}, 9227--9235\relax
\mciteBstWouldAddEndPuncttrue
\mciteSetBstMidEndSepPunct{\mcitedefaultmidpunct}
{\mcitedefaultendpunct}{\mcitedefaultseppunct}\relax
\EndOfBibitem
\bibitem[Cioslowski(1991)]{cioslowski:91pra}
Cioslowski,~J. Density-driven self-consistent-field method: Density-constrained
  correlation energies in the helium series. \emph{Phys. Rev. A} \textbf{1991},
  \emph{43}, 1223--1228\relax
\mciteBstWouldAddEndPuncttrue
\mciteSetBstMidEndSepPunct{\mcitedefaultmidpunct}
{\mcitedefaultendpunct}{\mcitedefaultseppunct}\relax
\EndOfBibitem
\bibitem[Valderrama \latin{et~al.}(1999)Valderrama, Lude{\~n}a, and
  Hinze]{valderrama:99jcp}
Valderrama,~E.; Lude{\~n}a,~E.~V.; Hinze,~J. {Assessment of dynamical and
  nondynamical correlation energy components for the beryllium-atom
  isoelectronic sequence}. \emph{J. Chem. Phys.} \textbf{1999}, \emph{110},
  2343--2353\relax
\mciteBstWouldAddEndPuncttrue
\mciteSetBstMidEndSepPunct{\mcitedefaultmidpunct}
{\mcitedefaultendpunct}{\mcitedefaultseppunct}\relax
\EndOfBibitem
\bibitem[Ramos-Cordoba \latin{et~al.}(2016)Ramos-Cordoba, Salvador, and
  Matito]{ramos-cordoba:16pccp}
Ramos-Cordoba,~E.; Salvador,~P.; Matito,~E. Separation of dynamic and
  nondynamic correlation. \emph{Phys. Chem. Chem. Phys.} \textbf{2016},
  \emph{18}, 24015--24023\relax
\mciteBstWouldAddEndPuncttrue
\mciteSetBstMidEndSepPunct{\mcitedefaultmidpunct}
{\mcitedefaultendpunct}{\mcitedefaultseppunct}\relax
\EndOfBibitem
\bibitem[L{\"o}wdin(1955)]{lowdin:55pr}
L{\"o}wdin,~P.-O. Quantum theory of many-particle systems. I. Physical
  interpretations by means of density matrices, natural spin-orbitals, and
  convergence problems in the method of configurational interaction.
  \emph{Phys. Rev.} \textbf{1955}, \emph{97}, 1474--1489\relax
\mciteBstWouldAddEndPuncttrue
\mciteSetBstMidEndSepPunct{\mcitedefaultmidpunct}
{\mcitedefaultendpunct}{\mcitedefaultseppunct}\relax
\EndOfBibitem
\bibitem[Smith~Jr(1967)]{smith:67tca}
Smith~Jr,~V.~H. \emph{Theor. Chim. Acta (Berlin)} \textbf{1967}, \emph{7},
  245\relax
\mciteBstWouldAddEndPuncttrue
\mciteSetBstMidEndSepPunct{\mcitedefaultmidpunct}
{\mcitedefaultendpunct}{\mcitedefaultseppunct}\relax
\EndOfBibitem
\bibitem[Kutzelnigg \latin{et~al.}(1968)Kutzelnigg, Del~Re, and
  Berthier]{kutzelnigg:68pr}
Kutzelnigg,~W.; Del~Re,~G.; Berthier,~G. Correlation coefficients for
  electronic wave functions. \emph{Phys. Rev.} \textbf{1968}, \emph{172},
  49\relax
\mciteBstWouldAddEndPuncttrue
\mciteSetBstMidEndSepPunct{\mcitedefaultmidpunct}
{\mcitedefaultendpunct}{\mcitedefaultseppunct}\relax
\EndOfBibitem
\bibitem[Lee \latin{et~al.}(1989)Lee, Rice, Scuseria, and Schaefer]{lee:89tca}
Lee,~T.~J.; Rice,~J.~E.; Scuseria,~G.~E.; Schaefer,~H.~F. A Diagnostic for
  Determining the Quality of Single-Reference Electron Correlation Methods.
  \emph{Theor. Chim. Acta (Berlin)} \textbf{1989}, \emph{75}, 81\relax
\mciteBstWouldAddEndPuncttrue
\mciteSetBstMidEndSepPunct{\mcitedefaultmidpunct}
{\mcitedefaultendpunct}{\mcitedefaultseppunct}\relax
\EndOfBibitem
\bibitem[Lee and Taylor(1989)Lee, and Taylor]{lee:89ijqc}
Lee,~T.~J.; Taylor,~P.~R. A Diagnostic for Determining the Quality of
  Single-Reference Electron Correlation Methods. \emph{Int. J. Quant. Chem.}
  \textbf{1989}, \emph{23}, 199--207\relax
\mciteBstWouldAddEndPuncttrue
\mciteSetBstMidEndSepPunct{\mcitedefaultmidpunct}
{\mcitedefaultendpunct}{\mcitedefaultseppunct}\relax
\EndOfBibitem
\bibitem[Janssen and Nielsen(1998)Janssen, and Nielsen]{janssen:98cpl}
Janssen,~C.~L.; Nielsen,~I. M.~B. New diagnostics for coupled-cluster and
  M{\o}ller-Plesset perturbation theory. \emph{Chem. Phys. Lett.}
  \textbf{1998}, \emph{290}, 423\relax
\mciteBstWouldAddEndPuncttrue
\mciteSetBstMidEndSepPunct{\mcitedefaultmidpunct}
{\mcitedefaultendpunct}{\mcitedefaultseppunct}\relax
\EndOfBibitem
\bibitem[Lee(2003)]{lee:03cpl}
Lee,~T.~J. Comparison of the T1 and D1 diagnostics for electronic structure
  theory: a new definition for the open-shell D1 diagnostic. \emph{Chem. Phys.
  Lett.} \textbf{2003}, \emph{372}, 362--367\relax
\mciteBstWouldAddEndPuncttrue
\mciteSetBstMidEndSepPunct{\mcitedefaultmidpunct}
{\mcitedefaultendpunct}{\mcitedefaultseppunct}\relax
\EndOfBibitem
\bibitem[Cioslowski(1992)]{cioslowski:92tca}
Cioslowski,~J. Differential density matrix overlap: an index for assessment of
  electron correlation in atoms and molecules. \emph{Theor. Chim. Acta
  (Berlin)} \textbf{1992}, \emph{81}, 319--327\relax
\mciteBstWouldAddEndPuncttrue
\mciteSetBstMidEndSepPunct{\mcitedefaultmidpunct}
{\mcitedefaultendpunct}{\mcitedefaultseppunct}\relax
\EndOfBibitem
\bibitem[Grimme and Hansen(2015)Grimme, and Hansen]{grimme:15ang}
Grimme,~S.; Hansen,~A. A Practicable Real-Space Measure and Visualization of
  Static Electron-Correlation Effects. \emph{Angew. Chem. Int. Ed. Engl.}
  \textbf{2015}, \emph{54}, 12308--12313\relax
\mciteBstWouldAddEndPuncttrue
\mciteSetBstMidEndSepPunct{\mcitedefaultmidpunct}
{\mcitedefaultendpunct}{\mcitedefaultseppunct}\relax
\EndOfBibitem
\bibitem[Raeber and Mazziotti(2015)Raeber, and Mazziotti]{raeber:15pra}
Raeber,~A.; Mazziotti,~D.~A. Large eigenvalue of the cumulant part of the
  two-electron reduced density matrix as a measure of off-diagonal long-range
  order. \emph{Phys. Rev. A} \textbf{2015}, \emph{92}, 052502\relax
\mciteBstWouldAddEndPuncttrue
\mciteSetBstMidEndSepPunct{\mcitedefaultmidpunct}
{\mcitedefaultendpunct}{\mcitedefaultseppunct}\relax
\EndOfBibitem
\bibitem[Ziesche(2000)]{ziesche:00the}
Ziesche,~P. On relations between correlation, fluctuation and localization.
  \emph{J. Mol. Struct. (Theochem)} \textbf{2000}, \emph{527}, 35--50\relax
\mciteBstWouldAddEndPuncttrue
\mciteSetBstMidEndSepPunct{\mcitedefaultmidpunct}
{\mcitedefaultendpunct}{\mcitedefaultseppunct}\relax
\EndOfBibitem
\bibitem[Boguslawski \latin{et~al.}(2012)Boguslawski, Tecmer, Legeza, and
  Reiher]{boguslawski:12jpcl}
Boguslawski,~K.; Tecmer,~P.; Legeza,~O.; Reiher,~M. Entanglement measures for
  single-and multireference correlation effects. \emph{J. Phys. Chem. Lett.}
  \textbf{2012}, \emph{3}, 3129--3135\relax
\mciteBstWouldAddEndPuncttrue
\mciteSetBstMidEndSepPunct{\mcitedefaultmidpunct}
{\mcitedefaultendpunct}{\mcitedefaultseppunct}\relax
\EndOfBibitem
\bibitem[Fogueri \latin{et~al.}(2013)Fogueri, Kozuch, Karton, and
  Martin]{fogueri:13tca}
Fogueri,~U.~R.; Kozuch,~S.; Karton,~A.; Martin,~J.~M. A simple DFT-based
  diagnostic for nondynamical correlation. \emph{Theor. Chem. Acc.}
  \textbf{2013}, \emph{132}, 1--9\relax
\mciteBstWouldAddEndPuncttrue
\mciteSetBstMidEndSepPunct{\mcitedefaultmidpunct}
{\mcitedefaultendpunct}{\mcitedefaultseppunct}\relax
\EndOfBibitem
\bibitem[Juh{\'a}sz and Mazziotti(2006)Juh{\'a}sz, and Mazziotti]{juhasz:06jcp}
Juh{\'a}sz,~T.; Mazziotti,~D.~A. The cumulant two-particle reduced density
  matrix as a measure of electron correlation and entanglement. \emph{J. Chem.
  Phys.} \textbf{2006}, \emph{125}, 174105\relax
\mciteBstWouldAddEndPuncttrue
\mciteSetBstMidEndSepPunct{\mcitedefaultmidpunct}
{\mcitedefaultendpunct}{\mcitedefaultseppunct}\relax
\EndOfBibitem
\bibitem[Skolnik and Mazziotti(2013)Skolnik, and Mazziotti]{skolnik:13pra}
Skolnik,~J.~T.; Mazziotti,~D.~A. Cumulant reduced density matrices as measures
  of statistical dependence and entanglement between electronic quantum domains
  with application to photosynthetic light harvesting. \emph{Phys. Rev. A}
  \textbf{2013}, \emph{88}, 032517\relax
\mciteBstWouldAddEndPuncttrue
\mciteSetBstMidEndSepPunct{\mcitedefaultmidpunct}
{\mcitedefaultendpunct}{\mcitedefaultseppunct}\relax
\EndOfBibitem
\bibitem[Weinert and Davenport(1992)Weinert, and Davenport]{weinert:92prb}
Weinert,~M.; Davenport,~J. Fractional occupations and density-functional
  energies and forces. \emph{Phys. Rev. B} \textbf{1992}, \emph{45},
  13709\relax
\mciteBstWouldAddEndPuncttrue
\mciteSetBstMidEndSepPunct{\mcitedefaultmidpunct}
{\mcitedefaultendpunct}{\mcitedefaultseppunct}\relax
\EndOfBibitem
\bibitem[Ramos-Cordoba \latin{et~al.}(2014)Ramos-Cordoba, Salvador, Piris, and
  Matito]{ramos-cordoba:14jcp}
Ramos-Cordoba,~E.; Salvador,~P.; Piris,~M.; Matito,~E. Two new constraints for
  the cumulant matrix. \emph{J. Chem. Phys.} \textbf{2014}, \emph{141},
  234101\relax
\mciteBstWouldAddEndPuncttrue
\mciteSetBstMidEndSepPunct{\mcitedefaultmidpunct}
{\mcitedefaultendpunct}{\mcitedefaultseppunct}\relax
\EndOfBibitem
\bibitem[Piris and Ugalde(2014)Piris, and Ugalde]{piris:14ijqc}
Piris,~M.; Ugalde,~J. Perspective on natural orbital functional theory.
  \emph{Int. J. Quant. Chem.} \textbf{2014}, \emph{114}, 1169--1175\relax
\mciteBstWouldAddEndPuncttrue
\mciteSetBstMidEndSepPunct{\mcitedefaultmidpunct}
{\mcitedefaultendpunct}{\mcitedefaultseppunct}\relax
\EndOfBibitem
\bibitem[Pernal and Giesbertz(2015)Pernal, and Giesbertz]{pernal:15tcc}
Pernal,~K.; Giesbertz,~K. J.~H. Reduced Density Matrix Functional Theory
  (RDMFT) and Linear Response Time-Dependent RDMFT (TD-RDMFT). \emph{Top. Curr.
  Chem.} \textbf{2015}, \emph{368}, 125\relax
\mciteBstWouldAddEndPuncttrue
\mciteSetBstMidEndSepPunct{\mcitedefaultmidpunct}
{\mcitedefaultendpunct}{\mcitedefaultseppunct}\relax
\EndOfBibitem
\bibitem[Cioslowski \latin{et~al.}(2015)Cioslowski, Piris, and
  Matito]{cioslowski:15jcp}
Cioslowski,~J.; Piris,~M.; Matito,~E. Robust validation of approximate 1-matrix
  functionals with few-electron harmonium atoms. \emph{J. Chem. Phys.}
  \textbf{2015}, \emph{143}, 214101\relax
\mciteBstWouldAddEndPuncttrue
\mciteSetBstMidEndSepPunct{\mcitedefaultmidpunct}
{\mcitedefaultendpunct}{\mcitedefaultseppunct}\relax
\EndOfBibitem
\bibitem[Perdew \latin{et~al.}(1982)Perdew, Parr, Levy, and
  Balduz~Jr]{perdew:82prl}
Perdew,~J.~P.; Parr,~R.~G.; Levy,~M.; Balduz~Jr,~J.~L. Density-functional
  theory for fractional particle number: derivative discontinuities of the
  energy. \emph{Phys. Rev. Lett.} \textbf{1982}, \emph{49}, 1691\relax
\mciteBstWouldAddEndPuncttrue
\mciteSetBstMidEndSepPunct{\mcitedefaultmidpunct}
{\mcitedefaultendpunct}{\mcitedefaultseppunct}\relax
\EndOfBibitem
\bibitem[Chai(2012)]{chai:12jcp}
Chai,~J.-D. Density functional theory with fractional orbital occupations.
  \emph{J. Chem. Phys.} \textbf{2012}, \emph{136}, 154104\relax
\mciteBstWouldAddEndPuncttrue
\mciteSetBstMidEndSepPunct{\mcitedefaultmidpunct}
{\mcitedefaultendpunct}{\mcitedefaultseppunct}\relax
\EndOfBibitem
\bibitem[Chai(2014)]{chai:14jcp}
Chai,~J.-D. Thermally-assisted-occupation density functional theory with
  generalized-gradient approximations. \emph{J. Chem. Phys.} \textbf{2014},
  \emph{140}, 18A521\relax
\mciteBstWouldAddEndPuncttrue
\mciteSetBstMidEndSepPunct{\mcitedefaultmidpunct}
{\mcitedefaultendpunct}{\mcitedefaultseppunct}\relax
\EndOfBibitem
\bibitem[Fromager(2015)]{fromager:15mp}
Fromager,~E. On the exact formulation of multi-configuration density-functional
  theory: electron density versus orbitals occupation. \emph{Molec. Phys.}
  \textbf{2015}, \emph{113}, 419--434\relax
\mciteBstWouldAddEndPuncttrue
\mciteSetBstMidEndSepPunct{\mcitedefaultmidpunct}
{\mcitedefaultendpunct}{\mcitedefaultseppunct}\relax
\EndOfBibitem
\bibitem[Gr{\"u}ning \latin{et~al.}(2003)Gr{\"u}ning, Gritsenko, and
  Baerends]{gruning:03jcp}
Gr{\"u}ning,~M.; Gritsenko,~O.; Baerends,~E. Exchange-correlation energy and
  potential as approximate functionals of occupied and virtual Kohn--Sham
  orbitals: Application to dissociating H2. \emph{J. Chem. Phys.}
  \textbf{2003}, \emph{118}, 7183--7192\relax
\mciteBstWouldAddEndPuncttrue
\mciteSetBstMidEndSepPunct{\mcitedefaultmidpunct}
{\mcitedefaultendpunct}{\mcitedefaultseppunct}\relax
\EndOfBibitem
\bibitem[Perdew \latin{et~al.}(1996)Perdew, Ernzerhof, and Burke]{perdew:96jcp}
Perdew,~J.~P.; Ernzerhof,~M.; Burke,~K. Rationale for mixing exact exchange
  with density functional approximations. \emph{J. Chem. Phys.} \textbf{1996},
  \emph{105}, 9982--9985\relax
\mciteBstWouldAddEndPuncttrue
\mciteSetBstMidEndSepPunct{\mcitedefaultmidpunct}
{\mcitedefaultendpunct}{\mcitedefaultseppunct}\relax
\EndOfBibitem
\bibitem[Jaramillo \latin{et~al.}(2003)Jaramillo, Scuseria, and
  Ernzerhof]{jaramillo:03jcp}
Jaramillo,~J.; Scuseria,~G.~E.; Ernzerhof,~M. Local hybrid functionals.
  \emph{J. Chem. Phys.} \textbf{2003}, \emph{118}, 1068--1073\relax
\mciteBstWouldAddEndPuncttrue
\mciteSetBstMidEndSepPunct{\mcitedefaultmidpunct}
{\mcitedefaultendpunct}{\mcitedefaultseppunct}\relax
\EndOfBibitem
\bibitem[Arbuznikov and Kaupp(2014)Arbuznikov, and Kaupp]{arbuznikov:14jcp}
Arbuznikov,~A.~V.; Kaupp,~M. Towards improved local hybrid functionals by
  calibration of exchange-energy densities. \emph{J. Chem. Phys.}
  \textbf{2014}, \emph{141}, 204101\relax
\mciteBstWouldAddEndPuncttrue
\mciteSetBstMidEndSepPunct{\mcitedefaultmidpunct}
{\mcitedefaultendpunct}{\mcitedefaultseppunct}\relax
\EndOfBibitem
\bibitem[Johnson(2014)]{johnson:14jcp}
Johnson,~E.~R. Local-hybrid functional based on the correlation length.
  \emph{J. Chem. Phys.} \textbf{2014}, \emph{141}, 124120\relax
\mciteBstWouldAddEndPuncttrue
\mciteSetBstMidEndSepPunct{\mcitedefaultmidpunct}
{\mcitedefaultendpunct}{\mcitedefaultseppunct}\relax
\EndOfBibitem
\bibitem[De~Silva and Corminboeuf(2015)De~Silva, and Corminboeuf]{silva:15jcp}
De~Silva,~P.; Corminboeuf,~C. Local hybrid functionals with orbital-free mixing
  functions and balanced elimination of self-interaction error. \emph{J. Chem.
  Phys.} \textbf{2015}, \emph{142}, 074112\relax
\mciteBstWouldAddEndPuncttrue
\mciteSetBstMidEndSepPunct{\mcitedefaultmidpunct}
{\mcitedefaultendpunct}{\mcitedefaultseppunct}\relax
\EndOfBibitem
\bibitem[Henderson \latin{et~al.}(2007)Henderson, Izmaylov, Scuseria, and
  Savin]{henderson:07jcp}
Henderson,~T.~M.; Izmaylov,~A.~F.; Scuseria,~G.~E.; Savin,~A. The importance of
  middle-range Hartree-Fock-type exchange for hybrid density functionals.
  \emph{J. Chem. Phys.} \textbf{2007}, \emph{127}, 221103\relax
\mciteBstWouldAddEndPuncttrue
\mciteSetBstMidEndSepPunct{\mcitedefaultmidpunct}
{\mcitedefaultendpunct}{\mcitedefaultseppunct}\relax
\EndOfBibitem
\bibitem[Henderson \latin{et~al.}(2008)Henderson, Izmaylov, Scuseria, and
  Savin]{henderson:08jctc}
Henderson,~T.~M.; Izmaylov,~A.~F.; Scuseria,~G.~E.; Savin,~A. Assessment of a
  middle-range hybrid functional. \emph{J. Chem. Theory Comput.} \textbf{2008},
  \emph{4}, 1254--1262\relax
\mciteBstWouldAddEndPuncttrue
\mciteSetBstMidEndSepPunct{\mcitedefaultmidpunct}
{\mcitedefaultendpunct}{\mcitedefaultseppunct}\relax
\EndOfBibitem
\bibitem[Janesko \latin{et~al.}(2009)Janesko, Henderson, and
  Scuseria]{janesko:09pccp}
Janesko,~B.~G.; Henderson,~T.~M.; Scuseria,~G.~E. Screened hybrid density
  functionals for solid-state chemistry and physics. \emph{Phys. Chem. Chem.
  Phys.} \textbf{2009}, \emph{11}, 443--454\relax
\mciteBstWouldAddEndPuncttrue
\mciteSetBstMidEndSepPunct{\mcitedefaultmidpunct}
{\mcitedefaultendpunct}{\mcitedefaultseppunct}\relax
\EndOfBibitem
\bibitem[Iikura \latin{et~al.}(2001)Iikura, Tsuneda, Yanai, and
  Hirao]{iikura:01jcp}
Iikura,~H.; Tsuneda,~T.; Yanai,~T.; Hirao,~K. A long-range correction scheme
  for generalized-gradient-approximation exchange functionals. \emph{J. Chem.
  Phys.} \textbf{2001}, \emph{115}, 3540--3544\relax
\mciteBstWouldAddEndPuncttrue
\mciteSetBstMidEndSepPunct{\mcitedefaultmidpunct}
{\mcitedefaultendpunct}{\mcitedefaultseppunct}\relax
\EndOfBibitem
\bibitem[Baer \latin{et~al.}(2009)Baer, Livshits, and Salzner]{baer:09arpc}
Baer,~R.; Livshits,~E.; Salzner,~U. Tuned range-separated hybrids in density
  functional theory. \emph{Ann. Rev. Phys. Chem.} \textbf{2009}, \emph{61},
  85\relax
\mciteBstWouldAddEndPuncttrue
\mciteSetBstMidEndSepPunct{\mcitedefaultmidpunct}
{\mcitedefaultendpunct}{\mcitedefaultseppunct}\relax
\EndOfBibitem
\bibitem[Henderson \latin{et~al.}(2008)Henderson, Janesko, and
  Scuseria]{henderson:08jpca}
Henderson,~T.~M.; Janesko,~B.~G.; Scuseria,~G.~E. Range Separation and Local
  Hybridization in Density Functional Theory. \emph{J. Phys. Chem. A}
  \textbf{2008}, \emph{112}, 12530--12542\relax
\mciteBstWouldAddEndPuncttrue
\mciteSetBstMidEndSepPunct{\mcitedefaultmidpunct}
{\mcitedefaultendpunct}{\mcitedefaultseppunct}\relax
\EndOfBibitem
\bibitem[Knowles and Handy(1989)Knowles, and Handy]{knowles:89cpc}
Knowles,~P.; Handy,~N. \emph{Comput. Phys. Commun.} \textbf{1989}, \emph{54},
  75\relax
\mciteBstWouldAddEndPuncttrue
\mciteSetBstMidEndSepPunct{\mcitedefaultmidpunct}
{\mcitedefaultendpunct}{\mcitedefaultseppunct}\relax
\EndOfBibitem
\bibitem[Frisch \latin{et~al.}()Frisch, Trucks, Schlegel, Scuseria, Robb,
  Cheeseman, Scalmani, Barone, Mennucci, Petersson, Nakatsuji, Caricato, Li,
  Hratchian, Izmaylov, Bloino, Zheng, Sonnenberg, Hada, Ehara, Toyota, Fukuda,
  Hasegawa, Ishida, Nakajima, Honda, Kitao, Nakai, Vreven, Montgomery, Peralta,
  Ogliaro, Bearpark, Heyd, Brothers, Kudin, Staroverov, Kobayashi, Normand,
  Raghavachari, Rendell, Burant, Iyengar, Tomasi, Cossi, Rega, Millam, Klene,
  Knox, Cross, Bakken, Adamo, Jaramillo, Gomperts, Stratmann, Yazyev, Austin,
  Cammi, Pomelli, Ochterski, Martin, Morokuma, Zakrzewski, Voth, Salvador,
  Dannenberg, Dapprich, Daniels, Farkas, Foresman, Ortiz, Cioslowski, and
  Fox]{g09}
Frisch,~M.~J.; Trucks,~G.~W.; Schlegel,~H.~B.; Scuseria,~G.~E.; Robb,~M.~A.;
  Cheeseman,~J.~R.; Scalmani,~G.; Barone,~V.; Mennucci,~B.; Petersson,~G.~A.;
  Nakatsuji,~H.; Caricato,~M.; Li,~X.; Hratchian,~H.~P.; Izmaylov,~A.~F.;
  Bloino,~J.; Zheng,~G.; Sonnenberg,~J.~L.; Hada,~M.; Ehara,~M.; Toyota,~K.;
  Fukuda,~R.; Hasegawa,~J.; Ishida,~M.; Nakajima,~T.; Honda,~Y.; Kitao,~O.;
  Nakai,~H.; Vreven,~T.; Montgomery,~J.~A.,~{Jr.}; Peralta,~J.~E.; Ogliaro,~F.;
  Bearpark,~M.; Heyd,~J.~J.; Brothers,~E.; Kudin,~K.~N.; Staroverov,~V.~N.;
  Kobayashi,~R.; Normand,~J.; Raghavachari,~K.; Rendell,~A.; Burant,~J.~C.;
  Iyengar,~S.~S.; Tomasi,~J.; Cossi,~M.; Rega,~N.; Millam,~J.~M.; Klene,~M.;
  Knox,~J.~E.; Cross,~J.~B.; Bakken,~V.; Adamo,~C.; Jaramillo,~J.;
  Gomperts,~R.; Stratmann,~R.~E.; Yazyev,~O.; Austin,~A.~J.; Cammi,~R.;
  Pomelli,~C.; Ochterski,~J.~W.; Martin,~R.~L.; Morokuma,~K.;
  Zakrzewski,~V.~G.; Voth,~G.~A.; Salvador,~P.; Dannenberg,~J.~J.;
  Dapprich,~S.; Daniels,~A.~D.; Farkas,~{\"O}.; Foresman,~J.~B.; Ortiz,~J.~V.;
  Cioslowski,~J.; Fox,~D.~J. Gaussian~09 {R}evision {D}.01. Gaussian Inc.
  Wallingford CT 2009\relax
\mciteBstWouldAddEndPuncttrue
\mciteSetBstMidEndSepPunct{\mcitedefaultmidpunct}
{\mcitedefaultendpunct}{\mcitedefaultseppunct}\relax
\EndOfBibitem
\bibitem[Matito \latin{et~al.}(2010)Matito, Cioslowski, and
  Vyboishchikov]{matito:10pccp}
Matito,~E.; Cioslowski,~J.; Vyboishchikov,~S.~F. Properties of harmonium atoms
  from FCI calculations: Calibration and benchmarks for the ground state of the
  two-electron species. \emph{Phys. Chem. Chem. Phys.} \textbf{2010},
  \emph{12}, 6712\relax
\mciteBstWouldAddEndPuncttrue
\mciteSetBstMidEndSepPunct{\mcitedefaultmidpunct}
{\mcitedefaultendpunct}{\mcitedefaultseppunct}\relax
\EndOfBibitem
\bibitem[Ramos-Cordoba \latin{et~al.}(2015)Ramos-Cordoba, Lopez, Piris, and
  Matito]{ramos-cordoba:15jcp}
Ramos-Cordoba,~E.; Lopez,~X.; Piris,~M.; Matito,~E. H$_4$: A challenging system
  for natural orbital functional approximations. \emph{J. Chem. Phys.}
  \textbf{2015}, \emph{143}, 164112\relax
\mciteBstWouldAddEndPuncttrue
\mciteSetBstMidEndSepPunct{\mcitedefaultmidpunct}
{\mcitedefaultendpunct}{\mcitedefaultseppunct}\relax
\EndOfBibitem
\bibitem[Bulik \latin{et~al.}(2015)Bulik, Henderson, and
  Scuseria]{bulik:15jctc}
Bulik,~I.~W.; Henderson,~T.~M.; Scuseria,~G.~E. Can single-reference coupled
  cluster theory describe static correlation? \emph{J. Chem. Theory Comput.}
  \textbf{2015}, \emph{11}, 3171--3179\relax
\mciteBstWouldAddEndPuncttrue
\mciteSetBstMidEndSepPunct{\mcitedefaultmidpunct}
{\mcitedefaultendpunct}{\mcitedefaultseppunct}\relax
\EndOfBibitem
\bibitem[Robinson and Knowles(2012)Robinson, and Knowles]{robinson:12jcp}
Robinson,~J.~B.; Knowles,~P.~J. Application of the quasi-variational coupled
  cluster method to the nonlinear optical properties of model hydrogen systems.
  \emph{J. Chem. Phys.} \textbf{2012}, \emph{137}, 054301\relax
\mciteBstWouldAddEndPuncttrue
\mciteSetBstMidEndSepPunct{\mcitedefaultmidpunct}
{\mcitedefaultendpunct}{\mcitedefaultseppunct}\relax
\EndOfBibitem
\bibitem[Van~Voorhis and Head-Gordon(2000)Van~Voorhis, and
  Head-Gordon]{vanvoorhis:00jcp}
Van~Voorhis,~T.; Head-Gordon,~M. Benchmark variational coupled cluster doubles
  results. \emph{J. Chem. Phys.} \textbf{2000}, \emph{113}, 8873--8879\relax
\mciteBstWouldAddEndPuncttrue
\mciteSetBstMidEndSepPunct{\mcitedefaultmidpunct}
{\mcitedefaultendpunct}{\mcitedefaultseppunct}\relax
\EndOfBibitem
\bibitem[Jankowski and Kowalski(1999)Jankowski, and Kowalski]{jankowski:99jcp}
Jankowski,~K.; Kowalski,~K. Physical and mathematical content of
  coupled-cluster equations. II. On the origin of irregular solutions and their
  elimination via symmetry adaptation. \emph{J. Chem. Phys.} \textbf{1999},
  \emph{110}, 9345--9352\relax
\mciteBstWouldAddEndPuncttrue
\mciteSetBstMidEndSepPunct{\mcitedefaultmidpunct}
{\mcitedefaultendpunct}{\mcitedefaultseppunct}\relax
\EndOfBibitem
\bibitem[Ramos-Cordoba \latin{et~al.}(2012)Ramos-Cordoba, Matito, Mayer, and
  Salvador]{ramos-cordoba:12jctc}
Ramos-Cordoba,~E.; Matito,~E.; Mayer,~I.; Salvador,~P. Toward a Unique
  Definition of the Local Spin. \emph{J. Chem. Theory Comput.} \textbf{2012},
  \emph{8}, 1270--1279\relax
\mciteBstWouldAddEndPuncttrue
\mciteSetBstMidEndSepPunct{\mcitedefaultmidpunct}
{\mcitedefaultendpunct}{\mcitedefaultseppunct}\relax
\EndOfBibitem
\bibitem[Ramos-Cordoba \latin{et~al.}(2012)Ramos-Cordoba, Matito, Salvador, and
  Mayer]{ramos-cordoba:12pccp}
Ramos-Cordoba,~E.; Matito,~E.; Salvador,~P.; Mayer,~I. Local spins: improved
  Hilbert-space analysis. \emph{Phys. Chem. Chem. Phys.} \textbf{2012},
  \emph{14}, 15291--15298\relax
\mciteBstWouldAddEndPuncttrue
\mciteSetBstMidEndSepPunct{\mcitedefaultmidpunct}
{\mcitedefaultendpunct}{\mcitedefaultseppunct}\relax
\EndOfBibitem
\bibitem[Ramos-Cordoba and Salvador(2014)Ramos-Cordoba, and
  Salvador]{ramos-cordoba:14pccp}
Ramos-Cordoba,~E.; Salvador,~P. Diradical character from the local spin
  analysis. \emph{Phys. Chem. Chem. Phys.} \textbf{2014}, \emph{16},
  9565--9571\relax
\mciteBstWouldAddEndPuncttrue
\mciteSetBstMidEndSepPunct{\mcitedefaultmidpunct}
{\mcitedefaultendpunct}{\mcitedefaultseppunct}\relax
\EndOfBibitem
\bibitem[Pulay and Hamilton(1988)Pulay, and Hamilton]{pulay:88jcp}
Pulay,~P.; Hamilton,~T.~P. UHF natural orbitals for defining and starting
  MC-SCF calculations. \emph{J. Chem. Phys.} \textbf{1988}, \emph{88},
  4926--4933\relax
\mciteBstWouldAddEndPuncttrue
\mciteSetBstMidEndSepPunct{\mcitedefaultmidpunct}
{\mcitedefaultendpunct}{\mcitedefaultseppunct}\relax
\EndOfBibitem
\bibitem[Harriman(1964)]{harriman:64jcp}
Harriman,~J.~E. Natural Expansion of the First-Order Density Matrix for a
  Spin-Projected Single Determinant. \emph{J. Chem. Phys.} \textbf{1964},
  \emph{40}, 2827--2839\relax
\mciteBstWouldAddEndPuncttrue
\mciteSetBstMidEndSepPunct{\mcitedefaultmidpunct}
{\mcitedefaultendpunct}{\mcitedefaultseppunct}\relax
\EndOfBibitem
\end{mcitethebibliography}
\end{document}